\documentclass[preprint,12pt, a4paper]{elsarticle}
\usepackage{amssymb}
\usepackage{hyperref}
\usepackage[table]{xcolor}
\usepackage[T1]{fontenc}

\newcommand{\mycomment}[1]{}

\journal{preprint}

\begin{document}
\renewcommand{\labelenumii}{\arabic{enumi}.\arabic{enumii}}

\begin{frontmatter}

\title{Mov-Avg: Codeless time series analysis using moving averages}


\author[label1]{Paweł Weichbroth}
\author[label2]{Jakub Buczkowski}
\address[label1]{Gdansk University of Technology, Faculty of Electronics, Telecommunications and Informatics, Department of Software Engineering, Narutowicza 11/12, pawel.weichbroth@pg.edu.pl}
\address[label2]{Gdansk University of Technology, s188894@student.pg.edu.pl}

\begin{abstract}
This paper introduces Mov-Avg, the Python software package for time series analysis that requires little computer programming experience from the user. The package allows the identification of trends, patterns, and the prediction of future events based on data collected over time. In this regard, the Mov-Avg implementation provides three indicators to apply, namely: Simple Moving Average, Weighted Moving Average and Exponential Moving Average. Due to its generic design, the Mov-Avg software package can be used in any field where the application of moving averages is valid. In general, the Mov-Avg library for time series analysis contributes to a better understanding of data-driven processes over time by taking advantage of moving averages in any way adapted to the research context.
\end{abstract}

\begin{keyword}
moving average \sep time series analysis \sep codeless

\end{keyword}

\end{frontmatter}

\section*{Code Metadata}
\label{}

\begin{table}[!htb]
\begin{tabular}{|l|p{6.5cm}|p{6.5cm}|}
\hline
C1 & Current code version & 1.0 \\
\hline
C2 & Permanent link to code/repository used for this code version & \url{https://github.com/kubix23/PyMovePlot} \\
\hline
C3  & Permanent link to Reproducible Capsule & –\\
\hline
C4 & Legal Code License   & GPL-3.0 license \\
\hline
C5 & Code versioning system used & Git/GitHub\\
\hline
C6 & Software code languages, tools, and services used & Python\\
\hline
C7 & Compilation requirements, operating environments \& dependencies & Python 3.10, pandas 2.2.2+, matplotlib 3.9.2+, pandas-datareader 0.10.0+, mplfinance 0.12.10b0, tkinter 8.6.13\\
\hline
C8 & If available Link to developer documentation/manual & -- \\
\hline
C9 & Support email for questions &  jakbuczkow@gmail.com\\
\hline
\end{tabular}
\caption{Code metadata}
\label{codeMetadata} 
\end{table}

\section{Motivation and significance}
Time series analysis is an umbrella term that encompasses a set of methods devoted for analyzing a sequence of data points collected over a period of time \cite{akaike2012practice, alqahtani2021deep}.
With the recent increased interest at present due to the abundance of available data \cite{yu2024common}, in particular moving averages (MAs) \cite{box1976time} have been applied in many different areas with specific goals namely \cite{li2024continuous, zhuang2024rethinking}.

The most prominent concerns financial and stock market analysis, including trend identification: MAs help identify upward or downward trends in stock prices or other financial assets \cite{nabipour2020predicting, thakkar2020predicting, rezaei2021stock}; MAs form the basis of many fundamental and technical indicators, while crossovers of short- and long-term MAs can signal buying or selling opportunities \cite{huang2020testing, ayala2021technical}.

In the face of global warming and climate change, MAs are used to analyze long-term trends in temperature, precipitation \cite{viola2010analysis, karl2000record}, and other climate-related data by smoothing seasonal variations in weather data, making it easier to identify patterns or anomalies in climate change \cite{chang2018moving}.

In the field of public health and epidemiology, MAs are used for trend detection to track disease outbreaks, infections, or hospital admissions by smoothing daily or weekly data \cite{pelletier2021trends, piccialli2021artificial, schaffer2021interrupted}. MAs help identify long-term trends in public health statistics, such as mortality rates or vaccination coverage, as part of public health surveillance \cite{alzahrani2020forecasting, medema2020implementation, kogan2021early}.

With the goal of identifying and analyzing the economy both globally and locally, economists typically apply MAs to time-series data to project future economic conditions, cyclical patterns, and short- and long-term trends \cite{dritsakis2018forecasting, stratimirovic2018analysis, gudmundsson2021forecasting}. In this view, MAs serve to smooth volatile economic data (such as GDP, prices, unemployment rates), helping to distinguish between typical market "noise" and actual trend reversals, as well as to identify long-term trends \cite{proietti2003forecasting, nyangarika2019oil, hua2022back}.

In the case of signal processing, MAs are used to reduce noise in signals (e.g., audio, video, or sensor data) to focus on relevant patterns or underlying signals \cite{li2020novel, rizhky2023qrs, maddipatla2024performance}, as well as for filtering, where MAs act as simple low-pass filters to remove high-frequency components \cite{kawala2020comparison, luo2020frequency}.

Interestingly, MAs are also used in sports analytics, more specifically to track athlete performance metrics over time, helping coaches and analysts identify meaningful patterns and trends \cite{oliva2021differences, torres2022tracking}. On an individual level, in team sports, MAs help evaluate player performance and make informed decisions based on long-term data trends \cite{murray2017calculating, corbett2019change}.

In supply chain and inventory management, MAs are used to smooth sales data over time, enabling more accurate predictions of future demand \cite{aburto2007improved, khosroshahi2016bullwhip}. In the case of inventory control, MAs support established inventory reorder points by monitoring trends in product demand and consumption patterns.

Generally speaking, MAs are widely adopted in the areas of Data Science and Machine Learning, including two major applications, namely Time Series Forecasting and Feature Engineering. While the former involves using MAs in time series models to smooth historical data, identify trends, and improve forecast accuracy \cite{dinesh2021prediction, arunkumar2021forecasting}, then in the latter MAs are also extracted as "hidden" features in machine learning models, especially for predicting outcomes based on time-dependent data \cite{demir2019introducing, xiao2022research}.

Note that the research disciplines and specific applications listed above are not all of the areas in which MAs can provide relevant information. In summary, moving averages are valuable wherever there is a need to smooth data, identify trends, or reduce noise in time series data. 

To the best of our knowledge, three types of MAs are the most commonly used, viz: Simple Moving Average (SMA), Weighted Moving Average (WMA), or Exponential Moving Average (EMA), and are defined as follows:

\begin{itemize}
    \item \textbf{Simple moving average (SMA)}, an arithmetic average of a specified length calculated from \textit{n} historical data sets. 
    \begin{equation}
        SMA=\frac{a_{1}+a_{2}+a_{3}+...+a_{n}}{n}    
    \end{equation}
     
    \item \textbf{Weighted moving average (WMA)}, the sum of \textit{n} historical data, each multiplied by a weight equal to the next term of the arithmetic sequence. 
    \begin{equation}
        WMA=\frac{na_{1}+(n-1))a_{2}+...+a_{n}}{n+(n-1)+...+1} = \frac{\sum_{i=1}^{n}\sum_{j=1}^{n}a_{n-i}}{\frac{n(n+1)}{2}}
    \end{equation}
     \\
    where:
    \begin{equation}
        a_{i}=\frac{2(n-i)}{n(n+1)}
    \end{equation}
        
    \item \textbf{Exponential Moving Average (EMA)}, the sum of \textit{n} historical data, each multiplied by a weight equal to the next term of the geometric sequence. 
    \begin{equation}
        EMA=\frac{a_{1}+(1-q)a_{2}+...+(1-q)^{n}a_{n}}{1+(1-q)+...+(1-q)^{n}} = a_{0}*q+SMA_{yesterday}*(1-q)
    \end{equation}
   
    where:
    \begin{equation}
        a_{i}=\frac{q^{i}}{\frac{1-q^{n}}{1-q}}
    \end{equation}
    and
    \begin{equation}
        q =\frac{2}{n+1}
    \end{equation}
    
\end{itemize}

The Mov-Avg library has been developed to provide a faster and easier way to analyze time series for users with little or no experience in computer programming. Due to its generic nature, Mov-Avg can be used in any application that would benefit from the above indicators. With data science research becoming increasingly popular, the Mov-Avg provides the necessary means to handle large amounts of data with little effort from the user. 

\section{Software description}
The library features a graphical interface where the provided data is displayed. It calculates moving averages for a selected column of the data and also plots them on the chart. The argument indicated by the user is marked with a vertical line. Next to the cursor, the data corresponding to a given date, along with the calculated averages, are displayed.

\subsection{Software architecture}
\begin{figure}[!htb]
    \centering
    \includegraphics[width=0.8\textwidth]{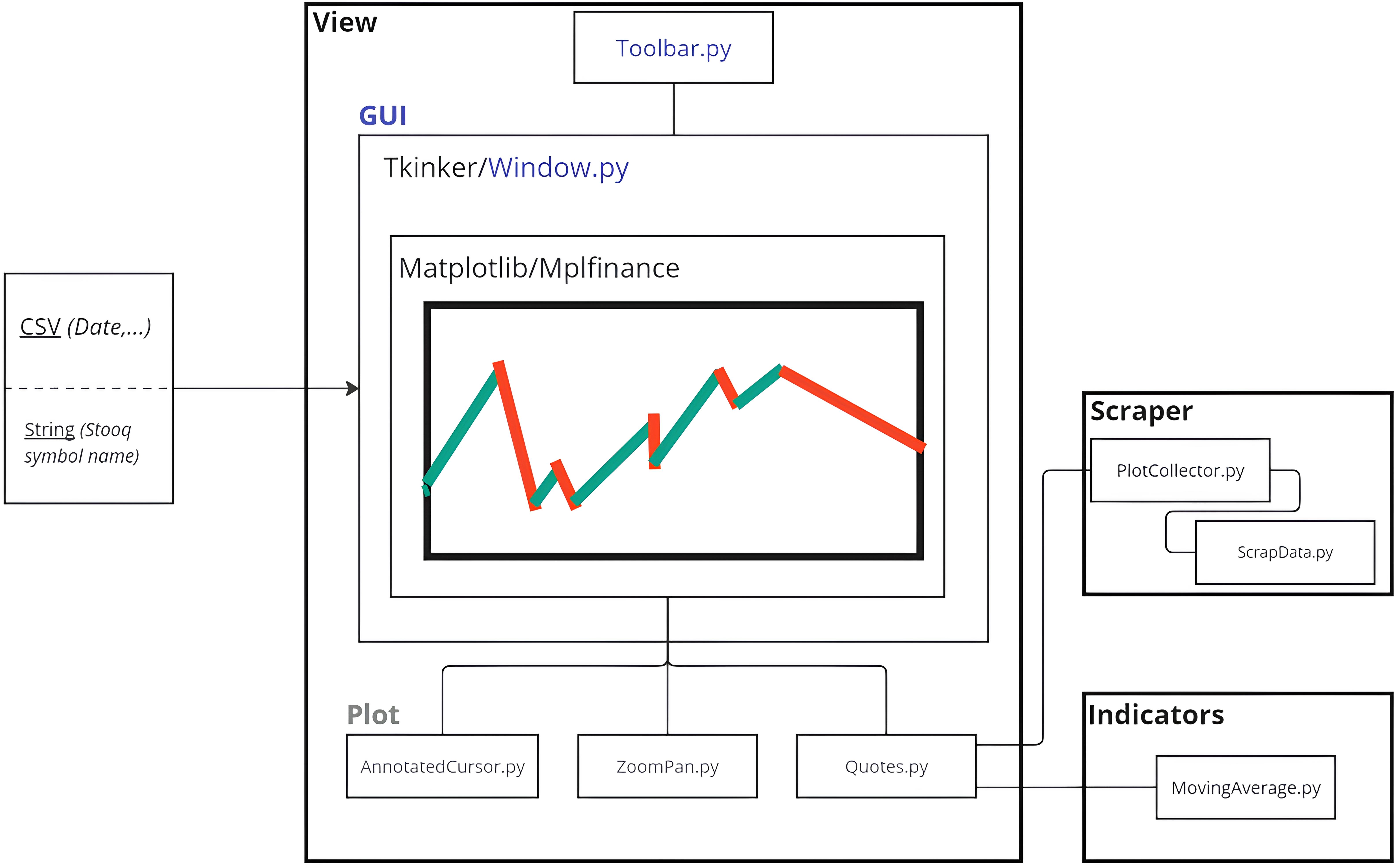}
    \caption{Library architecture showing graph functions}
    \label{figure1}
\end{figure}

The library is built on top of the Matplotlib library and its extension, Mplfinance. Two additional extensions have been created for it. The entire system is wrapped in a Tkinter backend. The Pandas-datareader library is also used to retrieve stock quotes from the Stooq website's API. A class is also created. The connection diagram is included in the image \ref{figure1}.

\begin{enumerate}
    \item \textit{Indicators}: The package responsible for the indicators displayed on the chart.
    \begin{itemize}
        \item \textit{Indicator.py}: Abstract class for stock market indices.
        \item \textit{MovingAverage.py}: Class that inherits from Indicator and which has a function that calculates one of three selected moving averages.
    \end{itemize}
    
    \item \textit{Scraper}: The package responsible for downloading data from the network.
    \begin{itemize}
        \item \textit{PlotCollector.py}: A class that acts as a data store for previously downloaded data from the network.
        \item \textit{ScrapData.py}: Class responsible for retrieving data with a given name.
    \end{itemize}

    \item \textit{View}: The package responsible for displaying and maintaining the chart.
    \begin{enumerate}
        \item \textit{GUI}: Package responsible for integrating Mplfinance with tkinker.
        \begin{itemize}
             \item \textit{Toolbar.py}: Class responsible for showing and hiding the default toolbar for matplotlib.
             \item \textit{Window.py}: The main class of the library, wraps matplotlib in a tkinker window.
        \end{itemize}

        \item \textit{Plot}: Package responsible for displaying data in Mplfinance and handling events related to it.
        \begin{itemize}
             \item \textit{AnnotatedCursor.py}: A class that creates a cursor that we use to move around the chart.
             \item \textit{Quotes.py}: The class creates an Mplfinance chart with selected indicators.
             \item \textit{ZoomPan.py}: Class that implements panning and zooming.
        \end{itemize}
    \end{enumerate}
\end{enumerate}

\subsection{Software functionalities}
\mycomment{
\textit{  Present the major functionalities of the software.}
}
Access to the library is provided by creating a \textit{Window} class and calling its \textit{showData} method with data in the form of a DataFrame (passed via the \textit{data=...} parameter) or a string representing a stock symbol (then passed via the \textit{name=...} parameter). Since \textit{Window} is a class that inherits from \textit{tkinker.Tk}, you can call the \textit{mainloop()} function on it, which is used to launch the program window.
The main functions of this library include the ability to display a time series provided by the user. Additionally, three moving averages of lengths chosen by the user are displayed. It is possible to change their lengths by passing the \textit{period=...} parameter in the  \textit{showData} function.

\section{Illustrative examples}

\subsection{Data}
The input data must have the first column as the Date. The subsequent columns represent the values you want to calculate. The entire system is built using Mplfinance, so the default names of the remaining columns should be in the format [Close, Open, High, Low]. This allows us to use display options other than just line charts. Table \ref{exampleData} presents sample data.
If you use column names other than the ones mentioned above (as long as the first column is the date), upon loading the data, the first column with data will be automatically duplicated into the required format.\\
Another option for loading data is to use the internet API and provide the name in the format like ticker symbol + "." + short country name e.g. "11B.PL", "EBAY.US". In this case, the program will use an internet connection to automatically retrieve the necessary data, store it in memory, and display it. 

\begin{table}[!hbt]
\caption{A fragment of the input data used to calculate the averages shown in Figure~\ref{figure2}.}
\label{exampleData}
\centering
\small
\begin{tabular}{|l|l|l|l|l|l|}
\hline
\textbf{Date}       & \textbf{Open}  & \textbf{High}  & \textbf{Low}   & \textbf{Close} & \textbf{Volume}  \\ \hline
04.02.2021 & 23.0  & 23.3  & 22.2  & 22.65 & 34676   \\ \hline
05.02.2021 & 22.85 & 23.3  & 22.05 & 22.1  & 31980   \\ \hline
08.02.2021 & 22.3  & 23.2  & 22.0  & 22.9  & 33666   \\ \hline
09.02.2021 & 22.9  & 22.9  & 22.0  & 22.0  & 27038   \\ \hline
10.02.2021 & 22.3  & 22.7  & 21.45 & 22.35 & 34944   \\ \hline
11.02.2021 & 22.6  & 22.7  & 22.15 & 22.7  & 14612   \\ \hline
12.02.2021 & 22.7  & 22.7  & 22.1  & 22.4  & 9663    \\ \hline
15.02.2021 & 22.4  & 22.8  & 22.05 & 22.6  & 15885   \\ \hline
16.02.2021 & 22.7  & 23.7  & 22.35 & 22.6  & 47222   \\ \hline
17.02.2021 & 22.95 & 23.2  & 22.3  & 22.8  & 26376   \\ \hline
18.02.2021 & 22.9  & 25.15 & 22.7  & 24.0  & 270015  \\ \hline
19.02.2021 & 24.3  & 25.25 & 23.1  & 24.8  & 109933  \\ \hline
22.02.2021 & 24.8  & 28.0  & 24.5  & 27.4  & 188952  \\ \hline
23.02.2021 & 25.0  & 26.2  & 24.0  & 25.2  & 182870  \\ \hline
24.02.2021 & 25.5  & 27.35 & 25.5  & 26.8  & 91654   \\ \hline
25.02.2021 & 27.05 & 28.55 & 26.9  & 27.9  & 85644   \\ \hline
26.02.2021 & 28.0  & 29.3  & 27.05 & 28.9  & 116575  \\ \hline
01.03.2021 & 30.0  & 32.8  & 30.0  & 32.65 & 235128  \\ \hline
02.03.2021 & 32.7  & 33.0  & 29.45 & 31.3  & 150421  \\ \hline
03.03.2021 & 45.0  & 59.2  & 41.2  & 59.2  & 1693818 \\ \hline
04.03.2021 & 74.5  & 98.0  & 59.2  & 59.2  & 2991466 \\ \hline
05.03.2021 & 61.0  & 70.9  & 60.4  & 63.0  & 999225  \\ \hline
08.03.2021 & 65.0  & 73.0  & 64.0  & 73.0  & 642558  \\ \hline
09.03.2021 & 73.0  & 88.3  & 72.0  & 87.4  & 1520660 \\ \hline
10.03.2021 & 81.0  & 89.0  & 75.6  & 79.3  & 858042  \\ \hline
11.03.2021 & 81.0  & 86.4  & 74.0  & 83.0  & 611475  \\ \hline
12.03.2021 & 86.0  & 94.6  & 84.6  & 93.5  & 922530  \\ \hline
15.03.2021 & 95.0  & 126.2 & 93.6  & 116.4 & 1364338 \\ \hline
16.03.2021 & 116.4 & 126.0 & 106.6 & 112.6 & 922690  \\ \hline
\end{tabular}
\end{table}

\subsection{Display}
\begin{figure}[!htb]
    \centering
    \includegraphics[width=0.8\textwidth]{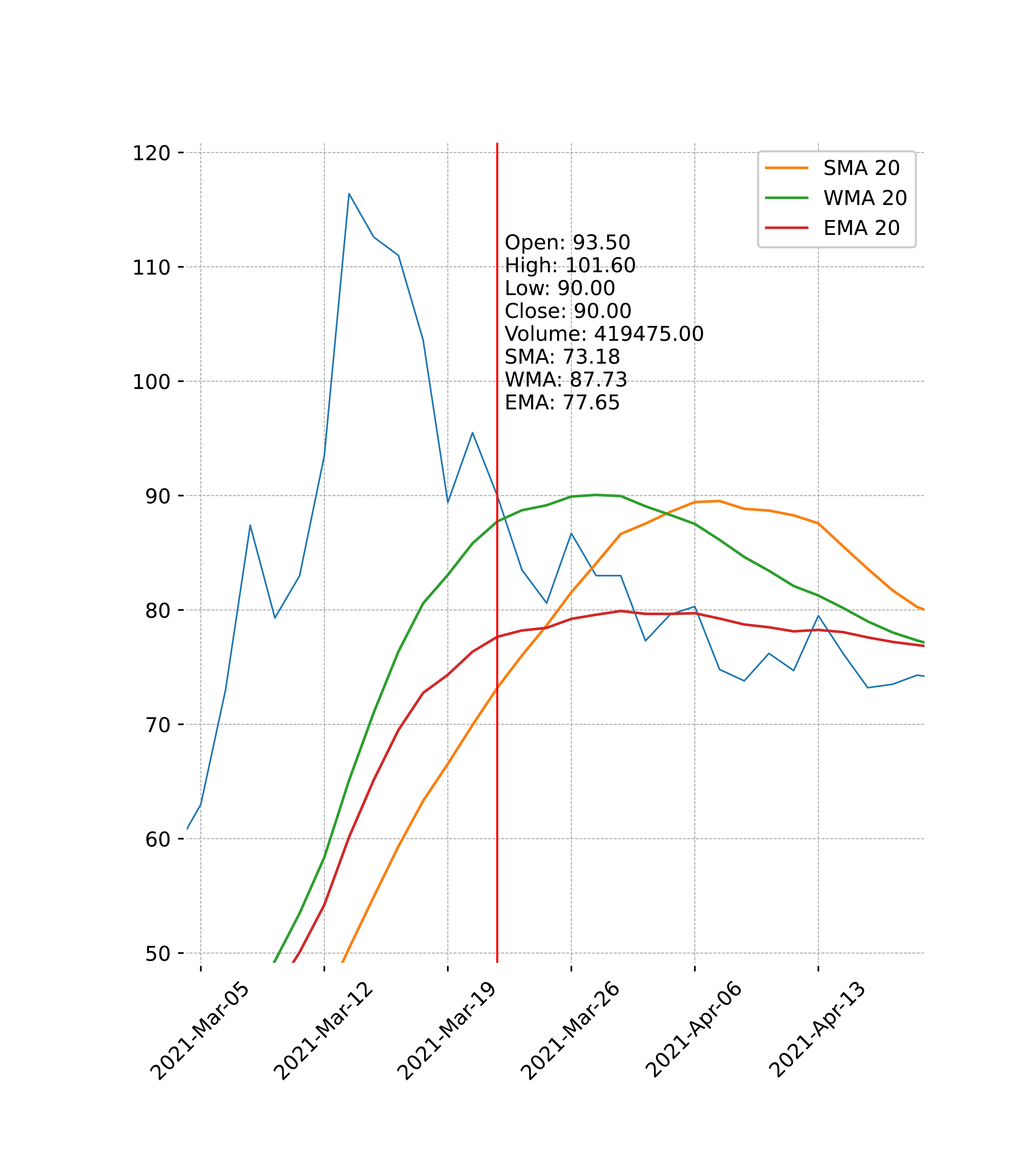}
    \caption{Example screen containing the displayed graph, with the cursor and data displayed as a line graph.}
    \label{figure2}
\end{figure}

On the user interface, data is displayed as a line chart by default, but using the parameter \textit{showData(plot\_type=...)}, you can specify one of the following types: 'candle', 'ohlc', 'line', 'renko', 'pnf', 'hollow\_and\_filled.\\
Based on the data to be displayed ('close' column), the following moving averages are calculated: SMA, WMA, and EMA. These averages are then overlaid on the chart. The length of these averages can be specified by the user (\textit{showData(period=int)}), and a legend is included on the chart, describing each of these averages along with their length.
The cursor on the chart responds to mouse movement, and next to it, the data for the currently selected date is displayed. The cursor automatically snaps to the nearest date and shows the data for it, with its position only changing vertically.\\
You can navigate the chart using the mouse. By clicking and dragging the chart in the desired direction, you can change its position, thus adjusting the segment of data you're viewing. The mouse scroll also allows zooming in and out. The default option scales the chart proportionally in both the \textit{X} and \textit{Y} axes, but by holding the \textit{Ctrl} key, you can adjust only the \textit{Y}-axis. The layout of the user interface is illustrated in Figure~\ref{figure2}.

\section{Impact}
Due to its generic design, the Mov-Avg software package can be used in any field where the application of moving averages is relevant, including finance and stock market analysis, economics, signal processing, supply chain and inventory management, weather and climate analysis, healthcare and epidemiology, sports analysis, to name a few.

In this view, Mov-Avg's design enables codeless development, initialization, and seamless monitoring of Time Series Analysis empirical research. This approach simplifies the mass execution of such experiments, allowing researchers to easily reproduce and evaluate numerous research results across different setups for self-specified time periods.

As a result, Mov-Avg increases the potential for research contributions in the field of time series forecasting. In industrial settings, it accelerates the deployment of accurate time series forecasting models in production by allowing domain experts to directly oversee the model development process. Its intuitive user interfaces match their domain expertise, even if their coding skills are limited. 

Overall, the Mov-Avg library for time series analysis contributes to a better understanding of dynamic processes over time by implementing moving averages tailored to the research context specified by the user. With our package installed, it is possible to efficiently perform data-driven research and quickly visualize the results with an easy-to-use view of specific data points. 

\section{Conclusions} 
In this paper we have presented and discussed the Mov-Avg library, an extensible and reliable open source software that allows codeless time series analysis based on the three basic indicators, including: Simple Moving Average, Weighted Moving Average, and Exponential Moving Average. By design, each indicator can be configured in any way by the user, taking into account the frequency of the data as well as the scope of the research. Possible future improvements to the software could include the implementation of other indicators to broaden the scope of the current analysis by supplying more information.

\section*{Declaration of competing interest}
The authors declare that they have no known competing financial interests or personal relationships that could have appeared to influence the work reported in this paper.

\bibliographystyle{elsarticle-num} 

\end{document}